An In-Class Discussion Activity on the Nature of Science and Intelligent Design

Brian C. Thomas, Assistant Professor
Department of Physics and Astronomy
Washburn University
1700 SW College Ave.
Topeka, KS 66621
brian.thomas@washburn.edu

In this paper I describe an in-class discussion activity aimed at helping elementary education majors in a physical science course think about issues surrounding the inclusion of "Intelligent Design" in public school science standards. I discuss the background instruction given, the content of the activity and some results from its use in class.

Introduction
Advocates of the philosophical/religious viewpoint known as "Intelligent Design" (ID) have been active across the country in recent years. The goal is to have their conclusions about the (supernatural) origins of biological diversity and physical structure in the universe taught in science classes as an "alternative" to the scientific theories of biological and physical evolution. In most cases, this activity centers around biology classes where biological evolution is taught. However, physics and astronomy are not immune, as cosmic evolution and the Big Bang theory are often explicitly opposed as well.[1] The APS, AAS and AAPT have all endorsed statements that ID and related ideologies should not be taught as science.[2] An additional resource that readers may find useful is the new journal *Evolution: Education and Outreach*.[3]

Intelligent Design is really a threat to the entire scientific enterprise, since it seeks to redefine the nature of science to include supernatural explanations. This redefinition is often subtle and the general public may see no reason to oppose such a definition, since they often have little knowledge about how science works. There has been significant discussion in the physics and astronomy community about how to teach non-science majors about what science is, partly in order to counteract the support that ID enjoys among the general public.[4]

The nature and process of science is a foundational topic in a course entitled "Physical Science for Elementary Educators" which I team-teach with a colleague in chemistry at Washburn University in Topeka, Kansas. I have developed and used in class a guided discussion activity intended to help the students think about the issues surrounding the status of ID as science and whether it should be included in state public school science standards.

Background Instruction
The first 2-3 days of class are spent covering background information on the nature and process of science. I carefully define terms such as "theory," "fact," "hypothesis," "model," "law," etc., and describe science as a process which seeks to understand the natural world through observation and experimentation and which excludes supernatural explanations.  I explain this restriction as a self-limitation, which is motivated by *principle* (the supernatural is not observable) and by *experience* (supernatural explanations rarely if ever lead to greater understating of natural processes).  I discuss pseudoscience in terms of breaking the "rules" of good science, and point out that such transgressions are at times committed by scientists as well.  I discuss the processes and structures we have in place to minimize such transgressions.  I carefully explain that the exclusion of the supernatural is truly a self-*limitation*, and not a statement of the non-existence of the supernatural. (That is, it is a methodological limitation, not a metaphysical one.)  I discuss the interaction of science with other areas of knowledge, including religion, drawing primarily from the viewpoint most famously described by Steven Jay Gould's "Nonoverlapping Magesteria."[5]

Science Standards in Kansas
Kansas is one state where science education standards have been a battleground.  The past decade or so has seen shifts back and forth as the makeup of the state school board has changed with elections. In 1999 and again in 2005 the Kansas State Board of Education rewrote science education standards to include language favorable to ID and critical of evolution.  The 2005 standards subtly redefined science itself, *excluding* science's self-imposed restriction to natural explanations, and implicitly allowing the interpretation of ID as valid science.  Elections (in 2000 and 2006) overturned the board's conservative majority and standards approved by the scientific community were reinstated.  Currently, the standards include a definition of science which corresponds to that generally agreed upon by scientists.

It is especially important in this ever-shifting and confusing environment that educators have a clear understanding of the nature of science and the nature of ID.  To that end, I developed a discussion activity to both help familiarize my students with the basic Kansas state science education standards, as well as to give them an opportunity to think about the status of ID in light of those standards and our discussions of the nature of science.

The Activity
The activity was used for the first time in August, 2007, with a class of 27 students (all elementary education majors). The students were given a partial copy of the standards[6] on the first day of class.  Included in this partial copy is the standards' definition of science, some discussion of the process of science and a statement regarding the respectful teaching of science.  The students were asked to read through this material on their own outside of class. The students were given the discussion activity on the third day of class, after instruction in the various terms, etc. regarding the nature and process of science. They were asked to work together (in groups of 3-4), to discuss their ideas and to write their thoughts down.

The activity itself consists of two parts. In Part 1, the students are asked to summarize the definition of science according to the standards. They are then asked to identify key words (and explain why these are important), and then to form their own definition particularly aimed at the grade level they plan to teach. Finally, they are asked whether they think there are any problems or inaccuracies with the standards' definition, and if so, how they would change the definition.

In Part 2, the students are given a simple definition of ID. That definition is as follows: "ID asserts that some features of the natural world cannot be explained as having come to be as they are through natural processes and that in those cases it is necessary to posit an external designer." I believe (and tell the students) that while they may encounter various incarnations of ID, this definition accurately represents the core of the viewpoint. I also make a strong point to the students that this activity is not intended to be a debate and that they are "not to attack, belittle or otherwise put-down your fellow students' ideas," and refer them to the section of the standards entitled "Teaching with Tolerance and Respect." This is done in order to maintain an environment where the students feel it is safe to be honest.

The discussion in Part 2 of the activity centers around two questions: 1) Is it correct to classify ID as science? and 2) Should ID viewpoints be taught in Kansas science classrooms? They are asked to write their answers, with explanations of their reasoning. Finally, they are asked to come to a consensus within their group and form a short "policy statement" regarding their position on these questions.

Results – Part 1
All the students found and recorded the definition presented in the standards that science is "a human activity of systematically seeking natural explanations for what we observe in the world around us." Many also added that science must be testable and repeatable. Key words which were chosen included "explanations," "observation," "natural," and "systematically." Grade-level definitions included a variety of simplifications or re-wordings, such as:
- "Science explains how the world works."
- "Science tells us why things happen." Or, "How and why things are, ex. 'Why is sky blue?'"
- "Science is understanding nature through observations based on different experiments."
- "How we explain what happens around us."
- "Using steps to explain the outside world."
- "A step by step method to discover more about our world."
- "Really figuring out how the world works, not just coming up with something."

The most interesting outcome in this part of the activity was comments by a few students (about 6 total) suggesting that the "natural" part of the definition was too limiting, or not appropriate. Some specific statements along these lines were:
- "Everything is not natural – there are accidents that are not natural."
- "Not everything has an explanation, some things just happen."
- "The word 'natural' may be a little too specific."
- "Not everything in science is natural, things will just happen."

One student specifically said that to write a better definition one should "just take out the word 'natural'." (This is particularly interesting considering the previous version of the standards did just that.)

These statements may indicate an inadequate understanding of "natural." For instance, "natural world" was misunderstood by a few students to only include the Earth, not the broader universe. They may also represent a desire to express belief in the reality of the supernatural (despite careful statements in class that methodological naturalism does not preclude such reality).

Results – Part 2
The questions posed generated active discussion among the students, which remained civil and calm. Most groups came to a statement along the lines of, "No, ID is not science. Maybe teach in another area as an idea about origins, etc., but not in science class." No group or individual stated that ID should be defined as science.

Many students (about 10) stated that ID *should* be taught, though "as a separate class." There were suggestions to include the viewpoint in philosophy or religion classes. Some modified this to state that ID is "not science, but should be mentioned as one viewpoint."

A few students (about 7) wrote that ID *should* be mentioned or taught in science classes, even though they had previously said it should *not* be defined as science. A few (about 4) couched this in terms of a "both sides" or "fairness" argument, which is a common tactic of advocates of ID. These statements tended to correlate with statements in Part 1 that the word "natural" was problematic in the definition of science.

Conclusions
Overall, this activity seems to have been a success – the students had a chance to engage with the science standards they will be teaching under (assuming no changes in the near future) and also had a chance to think about ID in that context. From the point of view of trying to limit the success of ID in making in-roads to public school science classrooms, the results are somewhat encouraging. Most students recognized that ID is not properly defined as science and should not be taught in science classrooms. However, there was a definite portion of the class (about 25%) that would certainly support such inclusion. Interestingly, this group seemed to see no contradiction between stating that ID is *not* science, but at the same time stating that it *should* be taught in science classrooms. This is a discouraging finding, since it indicates that arguments against ID being science, even if successful, may not effectively counteract all pressure to nevertheless include ID in science classrooms. Perhaps a more stark confrontation with such inconsistency would force a change in viewpoint, but this was not the point of the exercise, and may not be successful in any case, given the deep-seated nature of convictions that likely underlie such statements.

In conclusion, I would strongly encourage other instructors, especially those teaching courses for non-science majors and current or future educators, to not only clearly and directly discuss the nature of science in class,[7] but also to provide opportunities for students to think about and discuss together the nature of science and attempts to redefine science through ID or other movements. I believe that the locally relevant and intentionally non-confrontational nature of the activity described here is a constructive way to approach the topic, giving students a chance to think about the issues without putting them on the defensive. In future work I would like to further explore the contradictory viewpoints that arose and perhaps discover constructive ways for students to address these inconsistencies.

Acknowledgments
The author thanks an anonymous referee for helpful comments, especially pointing out the journal *Evolution: Education and Outreach*.

A reprint of this article can be found here:
http://www.stephenjaygould.org/library/gould_noma.html

6. Obtained from the Kansas Department of Education website:
http://www.ksde.org/Default.aspx?tabid=144

7. For an excellent discussion of the importance of defining terms, see: Helen Quinn, "Belief and knowledge—a plea about language," *Physics Today* 60(1), 8-9 (January 2007)